\documentclass{article}

\usepackage[preprint]{neurips_2025}

\usepackage[american]{babel}

\usepackage[utf8]{inputenc} 
\usepackage[T1]{fontenc}    
\usepackage{hyperref}       
\usepackage{url}            
\usepackage{booktabs}       
\usepackage{amsfonts}       
\usepackage{nicefrac}       
\usepackage{microtype}      
\usepackage{xcolor}         

\usepackage{mathtools} 
\usepackage{tikz} 

\usepackage{algorithm2e}
\usepackage{amsmath,amssymb}
\usepackage{mathtools}
\usepackage{subcaption}
\usepackage{wrapfig}
\usepackage{float}
\usepackage{graphicx}
\usepackage{enumitem}

\newcommand{\mc}{\mathcal}

\newtheorem{definition}{Definition}
\newtheorem{theorem}{Theorem}

\RestyleAlgo{ruled}

\allowdisplaybreaks

\title{Learning Recommender Mechanisms for Bayesian Stochastic Games}

\author{%
  Bengisu Guresti, Chongjie Zhang, and Yevgeniy Vorobeychik \\
  Washington University in St. Louis \\
  Saint Louis, MO \\
  \texttt{\{bengisu,chongjie,yvorobeychik\}@wustl.edu}
}

\begin{document}

\maketitle

\begin{abstract}
An important challenge in non-cooperative game theory is coordinating on a single (approximate) equilibrium from many possibilities—a challenge that becomes even more complex when players hold private information. 
Recommender mechanisms tackle this problem by recommending strategies to players based on their reported type profiles. 
A key consideration in such mechanisms is to ensure that players are incentivized to participate, report their private information truthfully, and follow the recommendations.
While previous work has focused on designing recommender mechanisms for one-shot and extensive-form games, these approaches cannot be effectively applied to stochastic games, particularly if we constrain recommendations to be Markov stationary policies.
To bridge this gap, we introduce a novel bi-level reinforcement learning approach for automatically designing recommender mechanisms in Bayesian stochastic games.
Our method produces a mechanism represented by a parametric function (such as a neural network), and is therefore highly efficient at execution time.
Experimental results on two repeated and two stochastic games demonstrate that our approach achieves social welfare levels competitive with cooperative multi-agent reinforcement learning baselines, while also providing significantly improved incentive properties.
\end{abstract}

\section{Introduction}

Equilibrium selection is a long-standing challenge in non-cooperative game theory.
It is particularly acute in repeated and stochastic games, in which even the set of pure-strategy equilibria can be quite large, as attested by the  folk theorems~\citep{fudenberg1991game}.
This fundamental coordination issue is further complicated if agents have private information (types)---arguably, the most typical scenario in practice---since the concomitant uncertainty can lead to further inefficiency and miscoordination.

A number of approaches have proposed addressing this issue by designing a \emph{recommender mechanism (RM)}, often referred to as a mediator,
which is a (possibly stochastic) mapping from joint player types to action recommendations with the property that players are incentivized to participate, report their types truthfully, and comply with the recommendations~\citep{kearns2014mechanism,myerson1982optimal}.
RMs have been successfully designed 
for one-shot~\citep{cummings2015privacy,ikegami2020simple,kearns2015robust,myerson1982optimal} and extensive-form games~\citep{zhang2022polynomial,zhang2024computing}.
However, no prior work has yet considered the design of recommender mechanisms in stochastic games (or the special case of repeated games).
This is a significant gap, as repeated and stochastic games have been a subject of an extensive body of research in \emph{multi-agent reinforcement learning (MARL)} when agents have potentially misaligned interests~\citep{bloembergen2015evolutionary,chakraborty2014multiagent,hu1998multiagent,conitzer2007awesome,powers2004new,sandholm2007perspectives}.
As highlighted by \citet{shoham2007if}, however, by focusing attention largely on the design of independent learning algorithms that converge to an equilibrium of a stage game, much of this literature does not truly tackle the underlying coordination problem.

We address this limitation by providing the first approach for designing recommender mechanisms in Bayesian stochastic games.
The mechanism we design takes a type profile of all players as an input and returns a profile of recommended \emph{policies}, one for each player.
The mechanics of this process is much like a routing app, such as Waze, in which agents report their origin and destination pairs (types), and which returns routes that each of them takes.
Our setup, however, accommodates an arbitrary Bayesian stochastic game that defines the environment.
We do not assume that this game is known in the sense of having an analytical description, but we do assume that a simulator for the game is available, as is commonly the case for multiagent learning problems considered in prior literature.
Our goal is to maximize social welfare,
subject to two incentive constraints.
The first of these is incentive compatibility (IC), which ensures that agents comply with the recommendations and report their types truthfully.
The second is individual rationality (IR), which incentivizes participation.

We impose three constraints on the nature of the mechanism: 1) recommendations are public, and 2) policies are Markov stationary and 3) deterministic.
The first means that the mechanism does not need to rely on a strong security implementation to make sure players do not learn information about one another's recommendations.
For example, if an approach recommends correlated strategies (as in prior work on extensive-form game mediators~\citep{zhang2024computing}), incentive properties depend critically on the ability of the designer to prevent any leaks of recommended actions among players.
The second is natural, since stochastic games always have Markov stationary equilibrium policies~\citep{fudenberg1991game}, and these are considerably simpler to implement than policies which require keeping track of arbitrarily long history of play.
Our focus on deterministic equilibria is limiting in general, but has a number of beneficial properties.
First, it considerably simplifies practical implementation (for example, avoiding reliance on pseudo-randomization that can lead to subtle errors~\citep{kelly2024zero}).
Second, a host of results suggest this restriction tends to be of lesser concern in complex scenarios.
For example, \citet{cartwright2009equilibrium} showed that in large (many-player) Bayesian games, any Bayesian equilibrium can be $\epsilon$-purified, while \citet{amiet2021pure} showed that when there are no ties, the probability that best response dynamics reaches a pure equilibrium quickly approaches 1 as the number of players grows.


We develop an approach for learning recommender mechanisms represented as a parametric function (e.g., a neural network) that casts the problem as \emph{bi-level reinforcement learning (BLRL)}. 
In this BLRL framework, the mechanism is learned as a meta-policy that takes type profiles as an input, while incentives are captured as lower-level RL subproblems involving individual agent deviations.
Our approach for solving this problem combines conventional RL loss terms for maximizing social welfare, with novel loss functions that capture IC and IR which are made differentiable with respect to the mechanism parameters through the use of straight-through gumbel-softmax \citep{gumbelsoftmax}, enabling an end-to-end gradient-based approach for mechanism learning.

We present experimental results on five domains. 
Two of these involve repeated bimatrix games, and the others represent more practically-relevant scenarios: a lane-changing game, and two congestion games.
Our results demonstrate that the proposed approach is highly effective in learning mechanisms that achieve high social welfare while limiting the incentives to misreport types or opt out.

In summary, we make the following contributions:
\begin{itemize}[leftmargin=*,itemsep=0pt,topsep=0pt]
    \item A framework, \emph{ReMBo}, for designing parametric \emph{policy \underline{re}commender \underline{m}echanisms for \underline{B}ayesian st\underline{o}chastic games},
    \item A \emph{bi-level reinforcement learning} approach for learning mechanism representations with high social welfare and strong incentives to opt in, report the types truthfully, and follow the recommendations, 
    \item Increased scalability using \emph{off-policy reinforcement learning} and a shared replay buffer for both training the mechanism and evaluating deviation incentives, and
    \item An \emph{experimental evaluation} showing that the proposed approach yields social welfare that is competitive with state-of-the-art MARL, while exhibiting significantly better incentive properties.
\end{itemize}
\section{Model}

\subsection{Environment}

We consider a Bayesian stochastic game among a set $I$ of $n$ agents who occupy the same stochastic dynamic environment.
Specifically, the environment is associated with a common and known state space $\mc{S} \subseteq \mathbb{R}^\ell$.
Each agent $i \in I$ has a finite action space $\mc{A}_i$ and a reward function $r_i(s,a;\theta_i)$, where $a = (a_1,\ldots,a_n)$ is an \emph{action profile} with $a_i \in \mc{A}_i$ for all $i$, and $\theta_i$ agent $i$'s \emph{type}.
While we assume that the reward function structure is common knowledge, agent types $\theta_i$ are \emph{private information} to each $i$ (that is, only $i$ knows it), inducing mutual uncertainty about reward functions.
We make the conventional assumption that $n$ is fixed and common knowledge.
Additionally, as is conventional in Bayesian games, we assume that there is a common knowledge prior distribution $\mc{P}_i$ over each agent's type $\theta_i \in \Theta_i$, where $\Theta_i$ is the finite space of possible types of agent $i$ and $\Theta = \Theta_1 \times \cdots \times \Theta_n$ the space of type profiles.
Let $\mc{A} = \mc{A}_1 \times \cdots \times \mc{A}_n$, and let the transition distribution be $\mc{T}:\mc{S} \times \mc{A} \times \mc{S} \rightarrow [0,1]$, where $\mc{T}(s'|s,a)$ is the probability that the next state is $s$ if the current state is $s$ and action profile $a \in \mc{A}$.
Finally, let $\rho$ be the initial state distribution, $H$ the time horizon of the stochastic game, and $\gamma \in (0,1)$ the temporal discount factor.

We restrict attention to deterministic Markov stationary policies for each agent.
Specifically, a policy $\pi_i(s,\theta_i)$ of an agent $i$ is a choice of $i$'s action $a_i \in \mc{A}_i$ in state $s$, when $i$'s type is $\theta_i$.
We use $\Pi_i$ to denote the space of possible policies for agent $i$.
Let $\pi = (\pi_1,\ldots,\pi_n)$ be a policy profile and $\Pi$ be the space of all policy profiles. 
We use $\theta_{-i}$ to denote a profile of types of agents other than $i$, and similarly $a_{-i}$ and $\pi_{-i}$ to refer to action and policy profiles, respectively, of all agents except $i$.
We define the ex-post game utility function of an agent $i$ as 
\[
    U_i(\pi,\theta )=\mathbb{E}\left[\sum_{t=0}^{H} \gamma^t r_i(s_t,a_t;\theta_i)|s_0 \sim \rho, a_t = \pi(s_t,\theta )\right].
\]
We define \emph{ex-post} social welfare as $U(\pi,\theta ) = \sum_i U_i(\pi,\theta )$.

\subsection{The Mechanism Design Problem}

We assume that the environment dynamics can be effectively represented by a simulator.
Our goal is to design a parametric policy recommender mechanism for Bayesian stochastic games (ReMBo).
ReMBo takes as input a vector of types $\theta = (\theta_1,\ldots,\theta_n)$ from all agents, and returns a policy profile $\pi$;
in other words, the mechanism is a mapping $\mc{M}:\Theta \rightarrow \Pi$.
Implicit in this is that we allow no payments as part of the mechanism.
We use $\mc{M}_i(\theta)$ to refer to a policy returned by the mechanism $\mc{M}$ for agent $i$, and similarly $\mc{M}_{-i}(\theta)$ will refer to policies for agents other than $i$.

A central considerations in the mechanism design problem involve eliminating incentives to game the system.
In our setting, three categories of incentives are particularly salient: ensuring that agents 1) report their types truthfully, 2) comply with the recommendations, and 3) are willing to participate (i.e., report their type at all).
Notably, the first two are entangled: incentives to lie about one's type may arise as a result of an agent's ability to subsequently fail to comply with the recommendation.
Consequently, we refer to the combination of (1) and (2) as \emph{incentive compatibility (IC)}.
We refer to the third consideration corresponds as \emph{individual rationality (IR)}.
Commonly, the consideration of these incentives involves either an entirely one-shot setting (e.g., a one-shot auction)~\citep{curry2023differentiable,dutting2019optimal,golowich2018deep}, or a series of repeated interactions between the designer (or mediator) and agents (as in dynamic mechanism design~\citep{bergemann2019dynamic,gerding2011online,hajiaghayi2007automated,parkes2003mdp}, as well as mediator design in extensive form games~\citep{zhang2024computing}).
Our case is different.
There is a single ``interaction'' phase between the agents and the designer, with agents reporting $\theta$ and the designer returning policies $\pi$.
Thereafter, agents interact \emph{dynamically} in the environment, with the designer no longer involved.\footnote{For example, \citet{zhang2022polynomial} and \citet{zhang2024computing} also assume that the interaction is dynamic, so that the designer need not reveal recommendations for any subgame not yet reached. In contrast, our setting involves recommending full policies to the agents.}
Next, we discuss IR and IC formally.

\noindent
\textbf{Individual Rationality } 
Our first consideration involves IR, requiring the mechanism to incentivize participation.
In our context, participation accounts for the difference between the utility that an agent $i$ would receive if they were to submit their type to the mechanism, compared to the utility they would obtain by keeping their type $\theta_i$ private and best responding to the profile of policies provided by the mechanism to all others.
Before we can proceed to formalize this, we note that our definition of the mechanism $\mc{M}$ does not directly accommodate the possibility that only $n-1$ agents submit their reports.
On the other hand, since $n$ is fixed and we only consider unilateral deviations, we need only to deal with the eventuality that only $n-1$ types are reported.
Consequently, we extend our definition of the mechanism as follows.
Let $\mc{M}(\theta) = \{\mc{M}^n(\theta),\{\mc{M}^{n-1}_{-i}(\theta_{-i})\}_{i \in I}\}$ be the mechanism that outputs a policy profile for every contingency whether all $n$ agents provide reports, or any one of them does not.

\begin{definition}
\label{defn:ir}
A mechanism $\mc{M}(\theta)$ 
is $\epsilon$-individually rational ($\epsilon$-IR) if for all $i \in I$,
    \begin{align}
        \label{E:IR}
        \mathbb{E}_\theta \left[\max_{\pi_{i}' \in \Pi_i}U_i(\pi_i',\mathcal{M}^{n-1}_{-i}(\theta_{-i}),\theta_i) - U_i(\mathcal{M}^n(\theta),\theta_i)\right] \le   \epsilon.
    \end{align}
\end{definition}

Henceforth, we abuse the notation slightly by using $\mc{M}(\theta)$ to refer to $\mc{M}^n(\theta)$ unless explicitly specified otherwise.

\noindent
\textbf{Incentive Compatibility } A typical notion of incentive compatibility in mechanism design is concerned with ensuring that no agent has an incentive to misreport their true type $\theta_i$.
A naive application of this concept in our setting would compare $i$'s utility from the policy $\mc{M}_i^n(\theta)$ (truthful reporting) with that from the $\mc{M}_i^n(\theta_i',\theta_{-i})$ (misreporting $i$'s type).
However, this fails to account for the ability of an agent to couple misreporting preferences with deviating from the policy returned by $\mc{M}^n$.
In other words, a benefit to $i$ from misreporting may be most pronounced when $i$ also leverages the lie to induce policies $\mc{M}_{-i}^n$ prescribed to other agents which are especially exploitable to $i$'s benefit.
This implies that we cannot decouple the issue of incentive compatibility from individual rationality in the sense of ensuring that each agent sticks to a policy provided by $\mc{M}^n$.
We therefore consider these in combination as aspects of IC.
Specifically, we consider approximate Bayes-Nash Incentive Compatibility (BNIC), which we formalize next.
\begin{definition}
\label{defn:ic}\label{E:bnic}
A mechanism $\mc{M}(\theta)$ is $\epsilon$-BNIC if for all $i \in I$,
\begin{align}
    \mathbb{E}_\theta\left[\max_{\pi_i' \in \Pi_i, \theta_{i}' \in \Theta_i}U_i(\pi_{i}', \mathcal{M}^n_{-i}(\theta_i',\theta_{-i}), \theta_i) 
     \vphantom{\max_{\pi_i' \in \Pi_i, \theta_{i}' \in \Theta_i}} 
     - U_i(\mathcal{M}^n(\theta), \theta_i)\right]
    \le \epsilon.
\end{align}
\end{definition}
Notably, our notion of $\epsilon$-BNIC couples misreports of types $\theta_i'$ and deviations from mechanism-prescribed policies.

\noindent
\textbf{Recommender Mechanism Design as an Optimization Problem } 
We assume that the designer's goal is to maximize \emph{ex ante} social welfare $\mathbb{E}_\theta[ U(\mc{M}(\theta),\theta)]$.
Since we admit no payments, there is no hope in our setting of achieving a ``first-best'' mechanism, that is, obtaining welfare-maximizing mechanisms while allowing no incentives to lie or deviate from prescribed policies~\citep{procaccia2013approximate}.
Consequently, it is useful to consider mechanism design as the following optimization problem:
\begin{equation}
\label{E:mdconstrained}
\begin{split}
& \max_{\mc{M}} \quad \mathbb{E}_\theta[ U(\mc{M}(\theta),\theta)]\\
& \textrm{s.t.} \quad \mathbb{E}_\theta\left[\max_{\pi_i' \in \Pi_i, \theta_{i}' \in \Theta_i}U_i(\pi_{i}', \mathcal{M}^n_{-i}(\theta_i',\theta_{-i}), \theta_i) 
\vphantom{\max_{\pi_i' \in \Pi_i, \theta_{i}' \in \Theta_i}} - U_i(\mathcal{M}^n(\theta), \theta_i)\right]
    \le \epsilon_1 \quad \forall i \in I 
    \\
& \quad \quad  \mathbb{E}_\theta \left[\max_{\pi_{i}' \in \Pi_i}U_i(\pi_i',\mathcal{M}^{n-1}_{-i}(\theta_{-i}),\theta_i) 
\vphantom{\max_{\pi_{i}' \in \Pi_i}} - U_i(\mathcal{M}^n(\theta),\theta_i)\right] \le   \epsilon_2 \quad \forall i \in I,
\end{split}
\end{equation}
where the first constraint corresponds to IC and the second to IR.
The following result follows directly.
\begin{theorem}
    Suppose $\mc{M}$ solves Problem~\eqref{E:mdconstrained}. Then participation, following the policy returned by $\mc{M}$, and reporting the true type $\theta_i$ for all agents $i$ constitutes an $\epsilon$-Bayes-Nash equilibrium, where $\epsilon= \max\{\epsilon_1,\epsilon_2\}$.
\end{theorem}

An important downside of this constrained optimization variant is that we need to specify a priori what the approximation bounds $\epsilon_1$ and $\epsilon_2$ are.
However, given that ideal is not achievable here, it is natural to instead aim to trade off the objective (social welfare) and incentives (IC and IR).
Consequently, we focus henceforth on the following unconstrained variation of the problem which constitutes its Lagrangian relaxation~\citep{boyd2004convex}:
\begin{equation}
\label{E:mdproblem}
\begin{split}
\max_{\mc{M}} \quad &\mathbb{E}_\theta\left[ U(\mc{M}(\theta),\theta) 
\right. \\
& - \lambda_1 \sum_{i \in I} \left(\max_{\pi_i' \in \Pi_i, \theta_{i}' \in \Theta_i}U_i(\pi_{i}', \mathcal{M}^n_{-i}(\theta_i',\theta_{-i}), \theta_i) 
\vphantom{\max_{\pi_i' \in \Pi_i, \theta_{i}' \in \Theta_i}} - U_i(\mathcal{M}^n(\theta), \theta_i)\right)\\
& -\lambda_{2}\sum_{i \in I}\left(\max_{\pi_{i}' \in \Pi_i}U_i(\pi_i',\mathcal{M}^{n-1}_{-i}(\theta_{-i}),\theta_i) 
\vphantom{U(\mc{M}(\theta),\theta)}\left.\vphantom{\max_{\pi_{i}' \in \Pi_i}} - U_i(\mathcal{M}^n(\theta),\theta_i)\right)\right]
\end{split}
\end{equation}
with $\lambda_1$ and $\lambda_2$ modulating the tradeoff between welfare, IC, and IR.

\section{Proposed Approach}

The question we address next is how to solve the mechanism design problem formalized in Equation~\eqref{E:mdproblem}.
This problem entails three challenges.
First, unlike past work on mechanism design~\citep{myerson1982optimal,kearns2015robust,zhang2024computing} including learning approaches~\citep{curry2023differentiable,dutting2019optimal,wang2024deep}, the utility function  $U_i(\pi,\theta)$ of each agent (and, consequently, the social welfare function $U(\pi,\theta)$ in the design problem) is not given explicitly.
Rather, it entails computation of total discounted reward over stochastic finite-length trajectories (e.g., state and control sequences in autonomous driving).
Second, we allow state and type space to be multi-dimensional (either finite or continuous), and consequently we must find a suitable mechanism representation.
Third, both the IR and IC terms entail solving inner maximization problems over policy and type spaces, a non-trivial proposition within an already non-trivial problem.

We address these challenges by leveraging neural network representations for 1) the mechanism $\mc{M}$ (for $\mc{M}^n$ and $\mc{M}^{n-1}_{-i}$), and 2) agent utility (value) functions $U_i$ for the three relevant contingencies: first, to evaluate the utility when all agents report types honestly and follow the prescribed policies, captured by $\mc{M}^n$, second, when agents deviate to play different policies and misreport types under $\mc{M}^n$, and third, when agents may not participate at all (i.e., under $\mc{M}^{n-1}$).
Furthermore, we make use of neural network representations for representing agent policies $\pi_i'$ that constitute deviations from $\mc{M}$.
Let $\phi$ aggregate all the parameters of the mechanism $\mc{M}$.
We let $\phi^n$ and $\phi^n_{-i}$ denote the parameters of $\mc{M}^n$ and $\mc{M}^n_{-i}$, respectively, and $\phi^{n-1}_{-i}$ to be the parameters of $\mc{M}^{n-1}_{-i}$.

Our first key observation is that the mechanism $\mc{M}$, with its constituent parts $\mc{M}^n$ and $\mc{M}^{n-1}_{-i}$, can be treated as \emph{policy profiles in the stochastic game parameterized by reported type profiles}.
Consequently, we can leverage the following connection between utility functions $U_i$ and agent value functions: for any policy profile $\pi$, $U_i(\pi,\theta_i) = \mathbb{E}_{s \sim \rho} [V_i(s,\pi,\theta_i))]$.
In other words, we can treat the mechanism design optimization problem in Equation~\eqref{E:mdproblem} as a \emph{bi-level reinforcement learning problem (RL)}, where the outer problem aims to learn the parameters of $\mc{M}$, while the inner problems in the IC and IR constraints aim, in turn, to learn near-optimal deviation policies $\pi_i'$ given a mechanism $\mc{M}$.
We rewrite Equation~\eqref{E:mdproblem} as the following problem:
\begin{equation}
\label{E:mdproblem_loss}
\min_{\phi} \mathbb{E}_\theta\left[\alpha_0 \mc{L}_{\mathit{RL}}(\phi^n) + \alpha_1 \mc{L}_{\mathit{IC}}(\phi^n,\theta) 
+\alpha_{2}\mc{L}_{\mathit{IR}}(\phi,\theta)\right],
\end{equation}
with $\alpha_0 + \alpha_1 + \alpha_2=1$ and all non-negative (essentially, rescaling the tradeoff coefficients in Equation~\eqref{E:mdproblem}), and where $\mc{L}_{\mathit{RL}}$ is a conventional RL loss, while $\mc{L}_{\mathit{IC}}$ and $\mc{L}_{\mathit{IR}}$ are the loss functions representing the IC and IR terms (which are, in turn, optimization problems).
These three loss terms represent three distinct subproblems: 1) maximize social welfare (represented by $\mc{L}_{\mathit{RL}}$), 2) minimize incentives to misreport types ($\mc{L}_{\mathit{IC}}$), and 3) maximize incentives to participate ($\mc{L}_{\mathit{IR}}$).
Next, we describe how we address each of these subproblems using gradient-based deep RL approaches.

\noindent
\textbf{Social Welfare Maximization }
In our setting, $\mc{L}_{\mathit{RL}}$ represents the goal of minimizing $-\mathbb{E}_{s \sim \rho}[\sum_i V_i(s,\mc{M}_\phi(\theta),\theta_i)]$, which we can view as the reinforcement learning problem in which the reward being maximized is the sum of individual agent rewards.
Consequently, we can plug in any conventional RL loss~\citep{dqn,levine2016end,lillicrap2015continuous}, and 
also leverage multiagent RL approaches~\citep{idqn1, maddpg,vdn,td3}.

\noindent
\textbf{Incentive Compatibility }
Let $\mc{L}_{\mathit{IC}} = \sum_{i \in I} \mc{L}_{\mathit{IC}}^i$.
Thus, we now focus on the IC loss term corresponding to a particular agent $i$, $\mc{L}_{\mathit{IC}}^i$, 
which represents the benefits to $i$ from misreporting type while also simultaneously choosing a different policy than the one returned by the mechanism.
Formally, this corresponds to the problem 
\begin{align*}
\max_{\pi_i'}\mathbb{E}_{s \sim \rho}[V_i(s,\pi_i',&\mc{M}^n_{\phi^n,-i}(\theta_i',\theta_{-i}),\theta_i)
-V_i(s,\mc{M}_{\phi^n}^n(\theta),\theta_i)].
\end{align*}
In principle, if we fix $\theta_i'$, this problem is itself an inner RL problem that computes the best response policy for agent $i$.
The challenge is that we wish to use gradient-based methods to learn the mechanism parameters $\phi$, but it is not straightforward how we can differentiate through the solution to this optimization problem.
One possible solution is to solve this inner RL problem for each update of $\phi$ and $\theta_i$ (thereby using an iterative gradient update scheme akin to what is done for generative adversarial networks~\citep{goodfellow2020generative}), but this is impractical because it entails complete RL runs for each gradient update of $\phi$.

We address this challenge by making use of learned action-value ($Q$) functions for agents in the context of the policies of other agents induced by the mechanism $\mc{M}$.
Specifically, let
\begin{align}
\label{E:lossic}
 \mc{L}_{\mathit{IC}}^i(\phi^n,\theta) = 
 \mathbb{E}_{s \sim \rho}\left[\sum_{\theta_{i}'} \sum_{a_i \in \mc{A}_i}\mathrm{ReLU}\left(Q_i(s, a_i, \mc{M}^n_{\phi^n,-i}(\theta_i',\theta_{-i}),\theta_i) 
 \vphantom{\sum_{\theta_{i}'} \sum_{a_i \in \mc{A}_i}}\vphantom{Q_i(s, a_i, \mc{M}^n_{\phi^n,-i}(\theta_i',\theta_{-i}),\theta_i)} - V_i(s,\mc{M}_{\phi^n}^n(\theta),\theta_i)\right)\right],
\end{align}
where $Q_i(\cdot)$ is the $Q$-function for agent $i$, which we represent as a neural network.

The next challenge is that neither $Q_i$ nor $V_i$ is differentiable with respect to $\phi$ when the set of actions $a_i$ is finite. 
We address this in two ways.
First, we encode the policy output of $\mc{M}$ as a vector (e.g., one-hot) encoding for each agent $i$.
Second, we relax maximization using the differentiable straight-through Gumbel-Softmax estimator \cite{gumbelsoftmax} for backpropagation.

The question that remains is how to estimate the functions $Q_i$ and $V_i$. We learn $Q_i$ by making use of conventional DQN loss while inferring the target actions $a'_{-i}$ of other agents from $\mc{M}^n_{\phi^n,-i}$ and taking the $\max_{a'_{i}}Q_i(s', a'_{i}, \mc{M}^n_{\phi^n,-i}(\theta),\theta_i)$. We learn $V_i$ using the conventional DQN loss by inferring the target actions $a'$ of all agents from $\mc{M}^n_{\phi^n}$ and using $V_i(s', \mc{M}^n_{\phi^n}(\theta),\theta_i)$ as the target. 
Note that since the mechanism $\mc{M}$ evolves through the training iterations, so do the estimated $Q_i$ and $V_i$.

\noindent
\textbf{Individual Rationality } \label{learnir}
As in the case of IC, $\mc{L}_{\mathit{IR}} = \sum_i \mc{L}^i_{\mathit{IR}}$, and so we focus here on the loss corresponding to agent $i$.
This corresponds to the problem 
\begin{align*}
\max_{\pi_i'}\mathbb{E}_{s \sim \rho}[V_i(s,\pi_i',&\mc{M}^{n-1}_{\phi^{n-1},-i}(\theta_{-i}),\theta_i)
-V_i(s,\mc{M}_{\phi^n}^n(\theta),\theta_i)].
\end{align*}
We again run into the same technical challenge as in the case of IC: needing to solve an inner RL problem.
We address it analogously as for IC, by leveraging the Q-learning-style updates.
Specifically, we use the following loss for the IR term:
\begin{align}
\label{E:lossir}
\mc{L}^i_{\mathit{IR}}(\phi,\theta) = 
\mathbb{E}_{s \sim \rho}\left[\sum_{a_{i}\in \mc{A}_i}\mathrm{ReLU} \left(Q_i^{n-1}(s, a_i, \mc{M}^{n-1}_{\phi^{n-1},-i}(\theta_{-i}),\theta) 
\vphantom{\sum_{a_{i}\in \mc{A}_i}}\vphantom{Q_i^{n-1}(s, a_i, \mc{M}^{n-1}_{\phi^{n-1},-i}(\theta_{-i}),\theta)} - V_i(s,\mc{M}_{\phi^n}^n(\theta),\theta_i)\right)\right],
\end{align}
where $Q_i^{n-1}$ is also represented as a neural network and learned iteratively (e.g., using DQN and similar techniques, as in the case of IC) in parallel with the mechanism $\phi$ during the full training phase. It is important to note that $Q_i^{n-1}$ and $Q_{-i}^{n-1}$ are learned as best responses to each other, even in our baselines where there is no IR loss, during this iterative learning to properly estimate their values since both depend on actions implied by the other as the best response. Although $Q_i^{n-1}$ estimates the utility of opting-out for agent $i$ and $Q_{-i}^{n-1}$ estimates the social welfare for rest of the agents which have no access to each other's policies $\pi'_{i}$ and $\mc{M}_{-i}^{n-1}$, centralized training for the best response policies has access to both $\pi'_{i}$ and $\mc{M}_{-i}^{n-1}$ to estimate the utilities accurately. Notably, $Q^{n-1}_i$ is not differentiable with respect to $\phi$ when the set of actions $a_i$ is discrete and finite. 
Here, we again run into the issue of differentiating $Q_i$ with respect to $\phi$; we address it in the same manner as in the case of IC above.

Inspecting the IR loss in Equation~\eqref{E:lossir} yields an interesting connection between our two-part mechanism that consists of $\mc{M}^n$ (when all participate) and $\mc{M}^{n-1}$ (when one agent opts out) and punishment strategies in dynamic games that help support desired equilibrium behavior.
Specifically, let us fix $\mc{M}^n$---the mechanism used on the (approximate) equilibrium path of play (when all agents follow the prescribed policies), and consider the mechanism $\mc{M}^{n-1}_{-i}$ designed for off-equilibrium paths when agent $i$ opts out.
Since parameters of $\mc{M}^{n-1}_{-i}$ are independent of the parameters of $\mc{M}^n$,  our IR loss effectively induces a worst-case deviation loss for the deviating agent.
In other words, the mechanism learns policies for all agents other than $i$ that maximally punish the deviator.
An important limitation of this approach is that while it is an approximate Bayes-Nash equilibrium (in the ex-ante sense), it is not in general a subgame perfect equilibrium, since the participating agents may not themselves be incentivized to follow the punishment phase.
However, it is natural to consider a Bayes-Nash equilibrium in our setting, since the policies produced by the mechanism would be represented in code, and so quite non-trivial to modify (and re-learn) in real-time.
Nevertheless, consideration of punishments that also account for incentives even in the punishment phase is an important direction for future work.

\noindent
\textbf{Putting Everything Together }
Putting everything together, we provide complete algorithms for both actor-critic-style (Algorithm~\ref{alg:cap}) and $Q$-learning-style methods (Algorithm~\ref{alg:cap2}), in the Appendix.
One important feature of our approach is that we can use a single shared replay buffer for the main recommender mechanism learning problem, as well as the IC and IR subproblems.
This enables us to reuse information obtained while learning different objectives (social welfare, non-compliance, or opting out) about the relationship between states, joint actions, and joint rewards, improving the efficacy of learning.
\section{Experiments}\label{experimentsection}

We evaluate our \emph{ReMBo} approach in three classes of Bayesian stochastic games domains: 1) repeated bimatrix games (specifically, \emph{Chicken} and \emph{Stag Hunt})~\citep{osborne1994course} which we generalize to incorporate incomplete information, 2) a lane-changing game (novel to this paper), and 3) two variants of congestion games~\citep{milchtaich1996congestion,Roughgarden2007AlgorithmicGT} in which player types correspond to origin-destination pairs.
We use three RL algorithmic approaches as both baselines and core building blocks for our approach: 
1) Multiagent TD3 (MA-TD3)~\citep{ddpg, maddpg, td3}, 2) VDN~\citep{vdn}, and 3) DQN with Team Reward (DQN-T)~\citep{idqn1, idqn2, idqn3}.
We use \emph{ReMBo-X} to refer to our ReMBo approach that leverages RL algorithm \emph{X} (e.g., \emph{ReMBo-MA-TD3}, etc).
We evaluate all approaches in terms of social welfare (Reward), IC violation (IC Loss, in the sense of Definition~\eqref{defn:ic}, with expectations estimated by sampling over type profiles and initial states, and deviations using standard RL techniques), and IR violation (in the sense of Definition~\eqref{defn:ir}).

\begin{figure*}[h]
    \centering
        \includegraphics[width=0.82\textwidth]{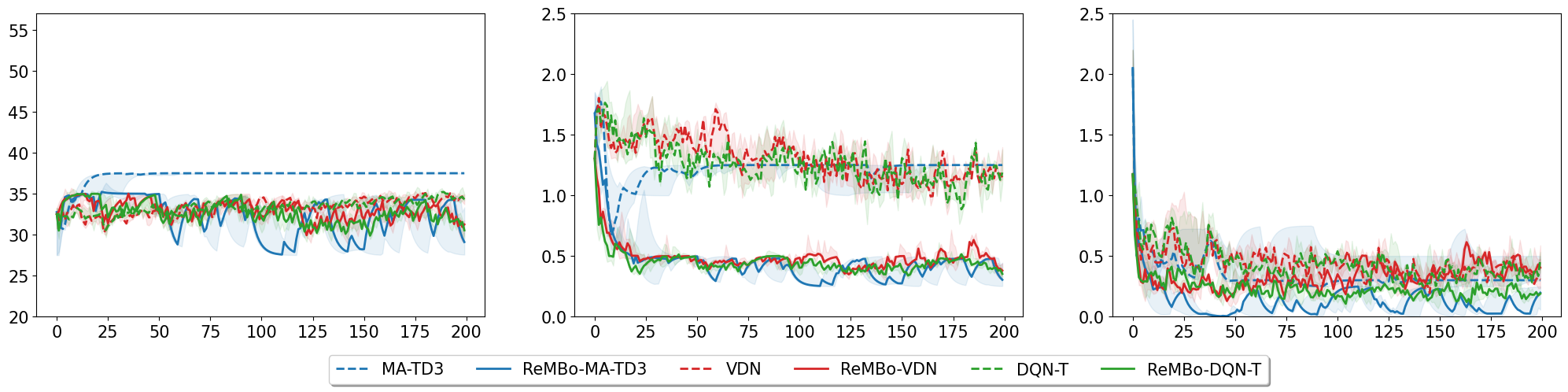}\\
        \includegraphics[width=0.82\textwidth]{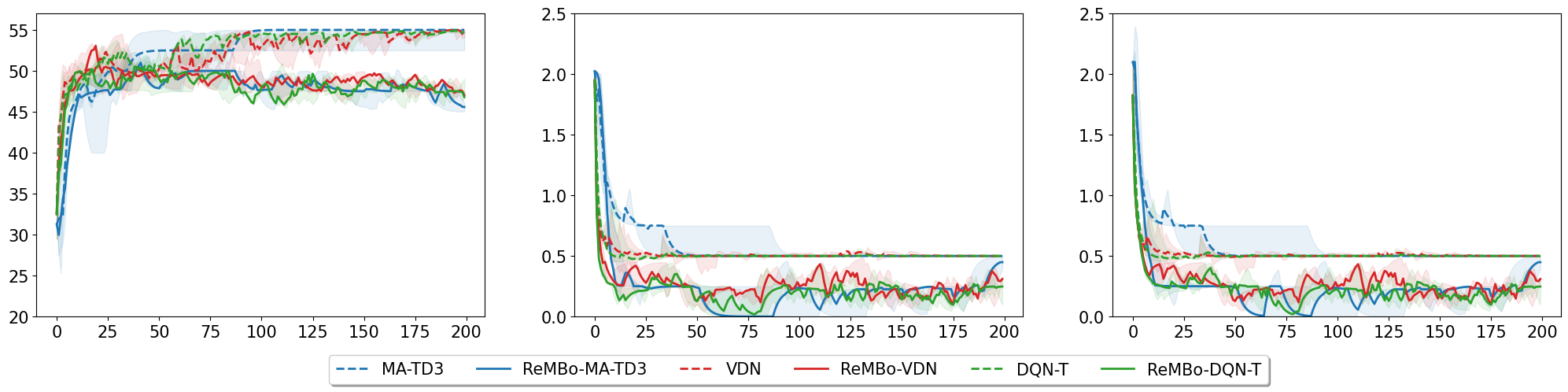}
    \caption{Results for \emph{Chicken} (top) and \emph{Stag Hunt} (bottom).  Left: social welfare, middle: IC deviation, right: IR deviation.}\label{fig:matrix}
\end{figure*}
\noindent
\textbf{Repeated Matrix Games }
We consider Bayesian variants of \emph{Chicken} and \emph{Stag Hunt}.
In the former, types are $\theta \in \{\text{Risk Averse (RA), Risk Taking (RT)}\}$, designed such that when both agents are RT, the game reduces to classic \emph{Chicken}.
For \emph{Stag Hunt}, types $\theta \in \{\text{Stag (S), Rabbit (R)} \}$ determine whether the agent prefers rabbit over stag or vice versa.
Full game structures are given in the Appendix, Tables~\ref{tab:chicken} and~\ref{tab:stag}.
Since these games are repeated, there is no meaningful state, and since the number of actions and types is small, we can compute IC and IR violations exactly.

The results for \emph{Chicken} and \emph{Stag Hunt} are provided in Figure~\ref{fig:matrix} (top and bottom, respectively).
In all cases, conventional RL approaches perform best in terms of social welfare (left plots), with MA-TD3 generally performing best.
The proposed \emph{ReMBo} approach, grafted onto each of the RL algorithms, tends to perform worse on this metric, although it is generally competitive.
In terms of incentives, however, the proposed \emph{ReMBo} approaches consistently outperform their baseline counterparts, in some cases significantly.
The performance improvement is particularly notable in \emph{Chicken} for IC, where the incentives to misreport types are several factors greater for the baselines than \emph{ReMBo}.

\begin{figure*}[h]
    \centering
        \includegraphics[width=0.82\textwidth]{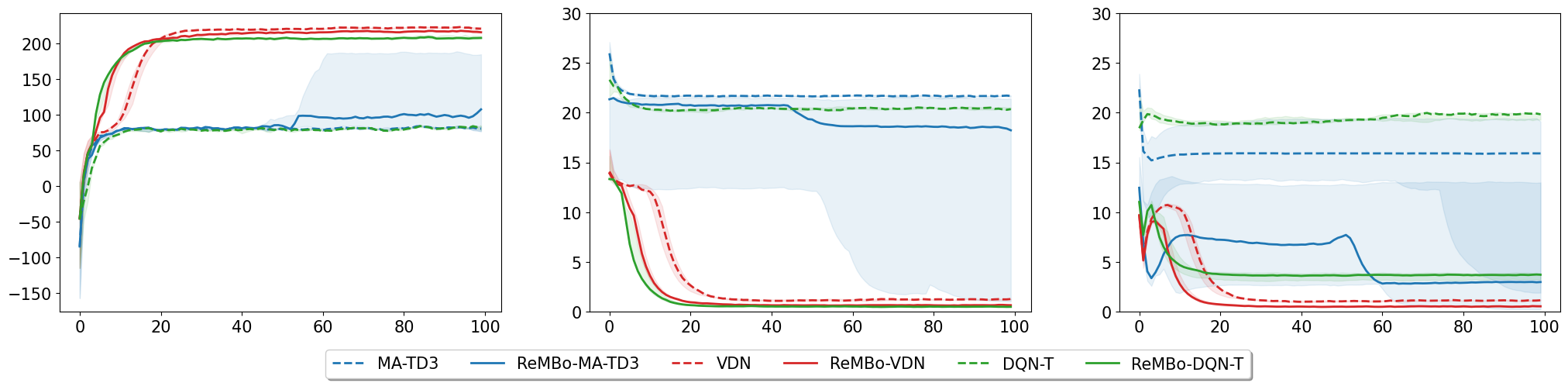}
        \includegraphics[width=0.82\textwidth]{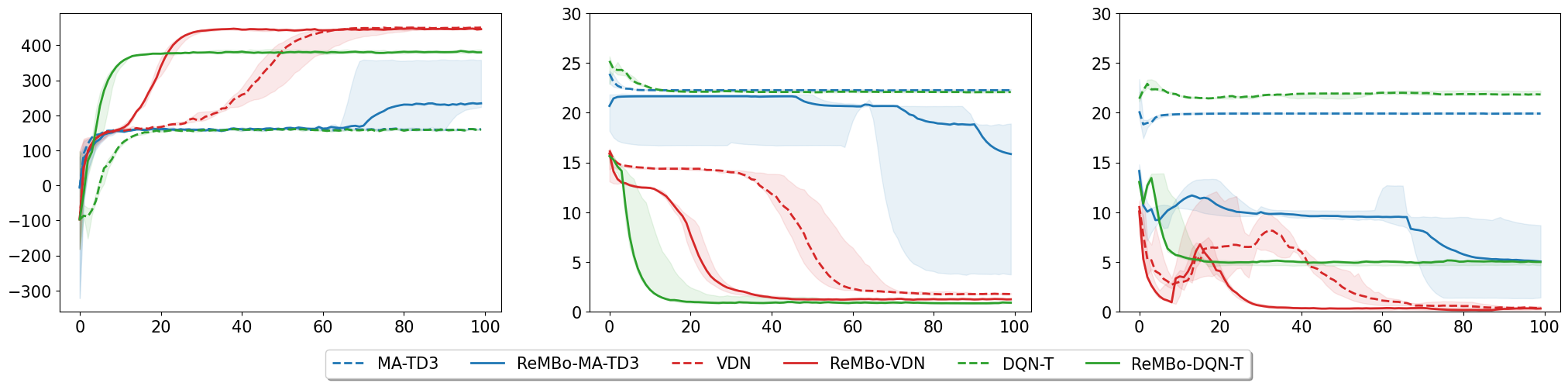}

    \caption{Results for the lane-changing game. 
 Top: 15 agents.  Bottom: 30 agents. 
 Left: social welfare, middle: IC deviation, and right: IR deviation.}\label{fig:lane_res}
\end{figure*}
\noindent
\textbf{Lane-Changing Game }
The next game we consider is a Bayesian \emph{lane-changing game} (Figure~\ref{fig:lane} in the Appendix), a stylized model of the lane changing dilemma drivers face.
The drivers enter to the game sequentially, and choose a lane when they enter.
Agents can have 2 target destinations $\theta \in \{\text{Target 1, Target 2}\}$, which are randomly assigned, and 2 opportunities to change lanes at fixed points which divide the roads into 3 parts. Agents are only affected by the congestion on the part of the road they are in or on the part of the road immediately in front of them. Lane 2 has $3\times$ the delay of Lane 1.
However if Lane 1 is too congested, Lane 2 could still lead to less delay. 
We consider two variants of this game: 1) 15 agents, 5 agents starting each step, and 2) 30 agents, 10 agents starting each step.




The results are shown in Figure \ref{fig:lane_res}.
The first surprising observation in this setting, which we will see again below, is that the proposed \emph{ReMBo} approaches tend to \emph{outperform their baseline RL counterparts in terms of social welfare}.
In all cases, they typically learn faster, and in most cases, they converge to higher social welfare, with \emph{VDN} being the sole exception.
We attribute this phenomenon to the \emph{ReMBo}'s improved ability to explore, stemming especially from the shared replay buffer that over time comes to include trajectories observed during deviation learning parts of the algorithm that are otherwise quite unlikely to be generated.
In terms of both IC and IR, \emph{ReMBo} consistently outperforms baselines, although \emph{VDN} ultimately converges (albeit, rather slowly) to be competitive.


\begin{figure*}[h]
    \centering
        \includegraphics[width=0.82\textwidth]{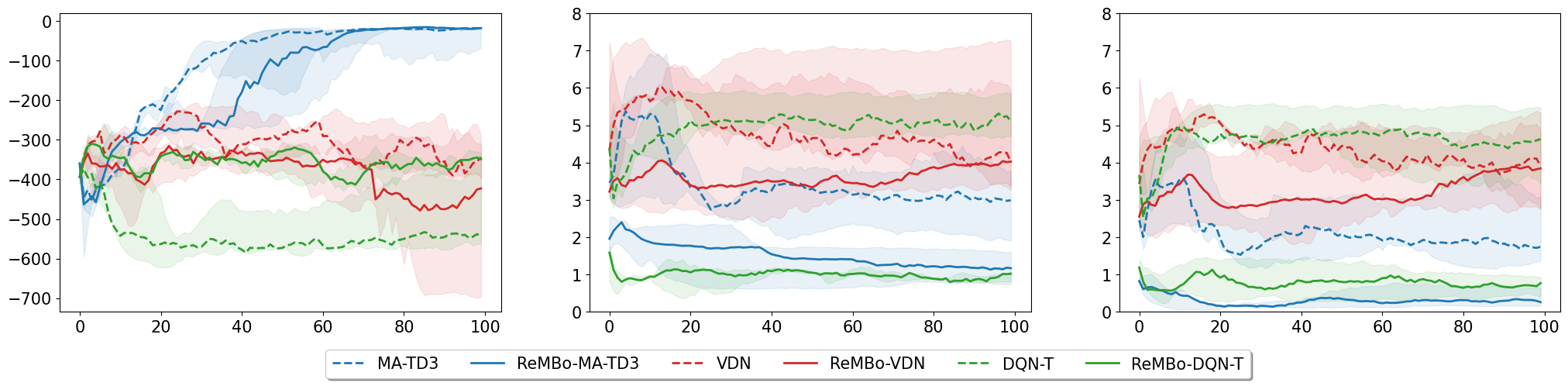}
        \includegraphics[width=0.82\textwidth]{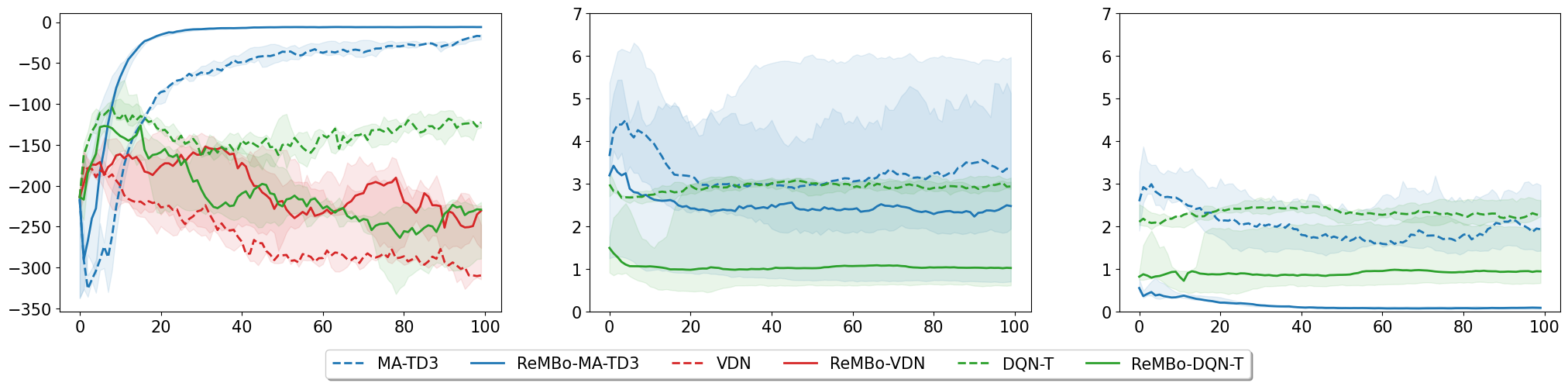}
    \caption{Results for congestion games. Top: 3-destination games.  Bottom: intersection games.  Left: social welfare, middle: IC deviation, and right: IR deviation.}
    \label{fig:congestion}
\end{figure*}
\noindent
\textbf{Congestion Games }
Finally, we consider two variants of congestion games: a \emph{3-destination game} (represented by the graph in Figure~\ref{fig:congmodels} (left) in the Appendix), in which all agents start at \emph{s} and have one of three destinations, which correspond to types, and an \emph{intersection game} (see Figure~\ref{fig:congmodels} (right) in the Appendix), in which agents have randomly chosen start and destination pairs.
The paths have congestion functions $c(x) = x$ and $c(x) = 1$, with $x$ denoting the fraction of agents using that path. 
For both games, we consider 10 agents.

The results are presented in Figure~\ref{fig:congestion}.
As in lane-changing games, we again observe several instances where \emph{ReMBo} actually outperforms the baseline in terms of social welfare, although there results on this metric are somewhat mixed.
Far more consistent are the IC and IR results, where \emph{ReMBo} in every case has considerably better incentive properties than the baseline it has been grafted onto.



\section{Related Work}
\label{relwork}
We contribute to the literature on recommender mechanism and mediator design~\citep{cummings2015privacy,ikegami2020simple,ivanov2023mediated,kearns2015robust,monderer2009strong,myerson1982optimal,zhang2022polynomial,zhang2024computing}.
In particular, our contribution is a model and approach for designing recommender mechanisms in Bayesian stochastic games, bridging an important gap in the literature that has considered either one-shot games~\citep{cummings2015privacy,ikegami2020simple,kearns2015robust,myerson1982optimal}, or interactive (dynamic) and correlated recommendations in extensive-form games~\citep{zhang2022polynomial,zhang2024computing}.
Notably, one-shot approaches cannot be effectively used for our setting because the normal-form representation of a Bayesian stochastic game which these require would itself be intractably large.

Our work is also situated at the intersection of multiagent learning and mechanism design.
In particular, an extensive multi-agent learning literature, however, addresses the issue of learning in non-cooperative settings.
However, this typically considers complete information games, with a focus on convergence to equilibria~\citep{bloembergen2015evolutionary,chakraborty2014multiagent,hu1998multiagent,conitzer2007awesome,powers2004new,sandholm2007perspectives}.
Also closely related is the work on \emph{mechanism design for multiagent planning}~\citep{van2008mechanism}, which tackles non-cooperative settings, but assumes a fully deterministic environment and allows payments.

Our work is also connected to the literature on dynamic and online mechanism design~\cite{bergemann2019dynamic,gerding2011online,hajiaghayi2007automated,parkes2003mdp}.
In this line of work, the mechanism and agents repeatedly interact (with agents possibly arriving and leaving), and the focus on predominantly on allocation settings, such as auctions.
In our setting, however, the mechanism is only involved in the process at the beginning, with compliance with recommendations a central consideration.
Moreover, we consider the problem of mechanism design without money, analogous to the broader literature such as market design~\citep{roth2018marketplaces,vulkan2013handbook} and facility location~\citep{chan2021mechanism,golowich2018deep,procaccia2013approximate}, where the lack of payments makes achieving perfect incentive alignment impossible in general~\citep{procaccia2013approximate}.
Finally, our work is situated within the field of automated mechanism design (AMD)~\citep{conitzer2003automated,vorobeychik2006empirical,vorobeychik2012constrained,zhang2021automated}, particularly the recent work that leverages neural network mechanism representations~\citep{dutting2019optimal,golowich2018deep,shen2018automated,wang2024deep}.
However, most prior work in AMD has considered one-shot settings, rather than dynamic stochastic games as we do.

\section{Conclusion}

We introduced a model and approach for (recommender) mechanism design in the context of Bayesian stochastic games.
In this setting, the designer takes player types as an input and provides policy recommendations to all players.
The key challenge lies in both achieving high social welfare, and incentivizing all agents to report types truthfully, follow the recommended policies, and participate in the mechanism (by reporting types).
Our proposed approach frames this as a bi-level RL problem, and grafts novel incentive compatibility (IC) and individual rationality (IR) losses into either actor-critic or $Q$-learning-style RL approaches.
One of the most surprising observations from our experiments is that the proposed approaches that aim to achieve IC and IR often converge much faster than conventional RL baselines \emph{in terms of social welfare}.
While this is not the case in repeated matrix games, we observe it in both lane-changing and congestion game experiments, suggesting that this is more than merely suboptimal hyperparameter choices.
We conjecture that this effect is due to improved exploration that the IC and IR constraints entail when coupled with a shared replay buffer, although this phenomenon warrants further investigation.


\newpage
\appendix

\section*{Appendix}
\section{Approach}\label{app:impl}

Here we provide the explicit loss functions to learn $Q_i$ and $V_i$ in Equations \ref{E:qibellman_loss}, and \ref{E:vibellman_loss} respectively. 
\begin{align}
\label{E:qibellman_loss}
 \mc{L}_{\mathit{Q}_{\psi^{Q}, i}}(\psi^{Q}, \mc{M}^n_{\phi^n,-i},\theta_i) =  \mathbb{E}_{s, a_i, a_{-i}, r_i, s' \sim \mc{D}}[(y_t - Q_{\psi^{Q}, i}(s, a_i, a_{-i},\theta_i))^2]
\end{align} where $y_t = r_i + \gamma \max\limits_{a'_i}Q_{\bar{\psi}^{Q}, i}(s, a'_i, \mc{M}^n_{\phi^n,-i}(s', \theta_{-i}),\theta_i)$.

\begin{align}
\label{E:vibellman_loss}
 \mc{L}_{\mathit{V}_{\psi^{V}, i}}(\psi^{V}, \mc{M}^n_{\phi^n},\theta_i) =  \mathbb{E}_{s, a, r_i, s' \sim \mc{D}}[(y_t - V_{\psi^{V}, i}(s, a,\theta_i))^2]
\end{align} where $y_t = r_i + \gamma V_{\bar{\psi}^{V}, i}(s, \mc{M}^n_{\phi^n}(s', \theta_{-i}),\theta_i)$.

We provide the pseudocode for the algorithm of MADDPG/TD3 Based ReMBo below.

\begin{algorithm}[H]
\caption{ReMBo - MADDPG/TD3 Based}\label{alg:cap}
    Initialize replay buffer $\mc{D}$\;
    Initialize $\mc{M}_{\phi^n}^n(\theta)$, $\mc{M}_{\phi^{n-1}}^{n-1}(\theta_{-i})$, and $\pi'_{\phi^{n-1}, i}(\theta_i)$ randomly\;
    Initialize $Q^n_{\psi^n}(\mc{M}^n_{\phi^n}(\theta),\theta)$, $Q^{n-1}_{\psi^{n-1}, -i}(\mc{M}^{n-1}_{\phi^{n-1}, -i}(\theta_{-i}),\theta)$ randomly\;
    Initialize $Q_{\psi^{Q}, i}(\mc{M}^n_{\phi^n, -i}(\theta),\theta_i)$, $V_{\psi^{V}, i}(\mc{M}^n_{\phi^n}(\theta),\theta_i)$, $Q^{n-1}_{\psi^{n-1}, i}(\pi'_{\phi', i}, \mc{M}^{n-1}_{\phi^{n-1}, -i}(\theta_{-i}),\theta_i)$ randomly\;
    Initialize target networks for all $\mc{M}$, $Q$, and $V$\;
    \For{$\text{episode} = 1..M$}{
        Get initial state $s_1$ and agent types $\theta$\;
        \For{$t = 1..T$}{
            Select actions $a_t$ with $\epsilon$ greedy exploration according to $\mc{M}_{\phi^n}^n(\theta)$\;
            Execute actions $a_t$ and get reward $r_t$ and next state $s_{t+1}$\;
            Store transition in $\mc{D}$\;
            Randomly sample a minibatch from $\mc{D}$\;
            Update $Q^{n-1}_{\psi^{n-1}, -i}(\mc{M}^{n-1}_{\phi^{n-1}}(\theta_{-i}),\theta)$ and $Q^{n-1}_{\psi^{n-1}, i}(\pi'_{\phi^{n-1}, i}, \mc{M}^{n-1}_{\phi^{n-1}, -i}(\theta_{-i}),\theta_i)$ minimizing $\mc{L}_{\mathit{RL}}$ for critic\;
            Update $\mc{M}_{\phi^{n-1}}^{n-1}(\theta_{-i})$ and $\pi'_{\phi^{n-1}, i}(\theta_i)$ minimizing $\mc{L}_{\mathit{RL}}$ for actor\;
            Update $Q^n_{\psi^{n}}(\mc{M}^n_{\phi^n}(\theta),\theta)$, $Q_{\psi^{Q}, i}(\mc{M}^n_{\phi^n,-i}(\theta),\theta_i)$, and $V_{\psi^{V}, i}(\mc{M}^n_{\phi^n}(\theta),\theta_i)$ minimizing $\mc{L}_{\mathit{RL}}$ for critic, Equation \ref{E:qibellman_loss} and Equation \ref{E:vibellman_loss} respectively\;
            Update $\mc{M}_{\phi^n}^n(\theta)$ and $\mc{M}_{\phi^{n-1}, -i}^{n-1}(\theta_{-i})$  minimizing the mechanism loss \ref{E:mdproblem_loss}\;
            Update target networks\;
        }
    }
\end{algorithm}

\newpage

We provide the pseudocode for the algorithm of VDN/DQN Based ReMBo below.

\begin{algorithm}[H]
\caption{ReMBo - VDN/DQN Based}\label{alg:cap2}
    Initialize replay buffer $\mc{D}$\;
    Initialize $Q^n_{\psi^n}(\theta)$, $Q^{n-1}_{\psi^{n-1}, -i}(\theta_{-i})$, and $Q^{n-1}_{\psi^{n-1}, i}(\theta_i)$ randomly\;
    Set $\mc{M}^n(\theta) \gets \mathop{\mathrm{argmax}}\limits_{a} Q^n_{\psi^n}(\theta)$\;
    Set $\mc{M}^{n-1}_{-i}(\theta_{-i}) \gets \mathop{\mathrm{argmax}}\limits_{a} Q^{n-1}_{\psi^{n-1}, -i}(\theta_{-i})$\;
    Set $\pi'_i(\theta_i) \gets \mathop{\mathrm{argmax}}\limits_{a} Q^{n-1}_{\psi^{n-1}, i}(\theta_i)$\;
    Initialize $Q_{\psi^Q, i}(\mc{M}^n_{\phi^n,-i}(\theta),\theta_i)$, $V_{\psi^V, i}(\mc{M}^n_{\phi^n}(\theta),\theta_i)$, $Q^{n-1}_{\psi^{n-1}, i}(\pi'_{\phi', i}, \mc{M}^{n-1}_{\phi^{n-1}}(\theta_{-i}),\theta_i)$ randomly\;
    Initialize target networks for all $Q$ and $V$\;
    \For{$\text{episode} = 1..M$}{
        Get initial state $s_1$ and agent types $\theta$\;
        \For{$t = 1..T$}{
            Select actions $a_t$ with $\epsilon$ greedy exploration according to $\mc{M}^n(\theta)$\;
            Execute actions $a_t$ and get reward $r_t$ and next state $s_t+1$\;
            Store transition in $\mc{D}$\;
            Randomly sample a minibatch from $\mc{D}$\;
            Update $Q^{n-1}_{\psi^{n-1}, -i}(\theta_{-i})$, $Q^{n-1}_{\psi^{n-1}, i}(\theta_i)$, and $Q^{n-1}_{\psi^{n-1}, i}(\pi'_{\phi', i}, \mc{M}^{n-1}_{\phi^{n-1}}(\theta_{-i}),\theta_i)$ minimizing $\mc{L}_{\mathit{RL}}$\;
            Update $Q_{\psi^Q, i}(\mc{M}^n_{\phi^n,-i}(\theta),\theta_i)$, and $V_{\psi^V, i}(\mc{M}^n_{\phi^n}(\theta),\theta_i)$ minimizing \ref{E:qibellman_loss} and \ref{E:vibellman_loss} respectively\;
            Update $Q^n_{\phi^n}(\theta)$, and $Q^{n-1}_{\psi^{n-1}, -i}(\theta_{-i})$ minimizing the mechanism loss \ref{E:mdproblem_loss}\;
            Update target networks\;
    }
}
\end{algorithm}

The implementations benefit from the repository jaxrl \cite{jaxrl}. The neural network architectures that are used in ReMBo are provided below. The vector notation $\vec{\pi}$ denotes the vector of action probabilities, $\vec{Q}$ denotes the vector of action values, and the scalar $Q$ denotes the scalar action-value.

\begin{figure}[H]
     \centering
     \begin{subfigure}[b]{0.25\textwidth}
         \centering
         \includegraphics[width=\textwidth]{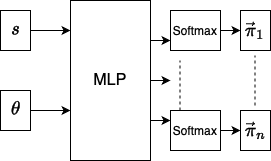}
         \caption{$\mc{M}_{\phi^n}^n(\theta)$}
         \label{fig:main_actor_mlp}
     \end{subfigure}
     \hfill
     \begin{subfigure}[b]{0.4\textwidth}
         \centering
         \includegraphics[width=\textwidth]{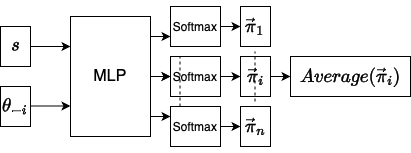}
         \caption{$\mc{M}_{\phi^{n-1}}^{n-1}(\theta_{-i})$}
         \label{fig:nopart_multi_actor_mlp}
     \end{subfigure}
     \hfill
     \begin{subfigure}[b]{0.3\textwidth}
         \centering
         \includegraphics[width=\textwidth]{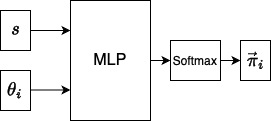}
         \caption{$\pi'_{\phi^{n-1}, i}(\theta_i)$}
         \label{fig:nopart_single_actor_mlp}
     \end{subfigure}
        \caption{(a) MLP actor architecture for the mechanism $\mc{M}_{\phi^n}^n(\theta)$, (b) MLP actor architecture for the mechanism $\mc{M}_{\phi^{n-1}}^{n-1}(\theta_{-i})$, (c) MLP actor architecture for the agent deviation policy $\pi'_{\phi^{n-1}, i}(\theta_i)$}
        \label{fig:mlp_actors}
\end{figure}

\begin{figure}[H]
     \centering
     \begin{subfigure}[b]{0.3\textwidth}
         \centering
         \includegraphics[width=\textwidth]{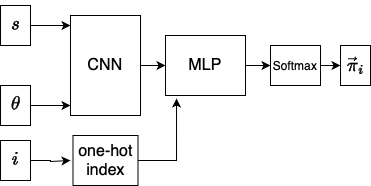}
         \caption{$\mc{M}_{\phi^n}^n(\theta)$}
         \label{fig:main_actor_mlp}
     \end{subfigure}
     \hfill
     \begin{subfigure}[b]{0.3\textwidth}
         \centering
         \includegraphics[width=\textwidth]{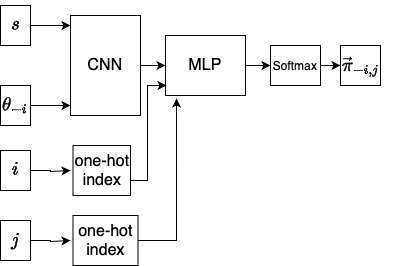}
         \caption{$\mc{M}_{\phi^{n-1}}^{n-1}(\theta_{-i})$}
         \label{fig:nopart_multi_actor_mlp}
     \end{subfigure}
     \hfill
     \begin{subfigure}[b]{0.3\textwidth}
         \centering
         \includegraphics[width=\textwidth]{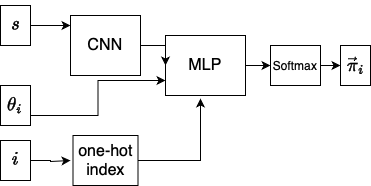}
         \caption{$\pi'_{\phi^{n-1}, i}(\theta_i)$}
         \label{fig:nopart_single_actor_mlp}
     \end{subfigure}
        \caption{(a) CNN actor architecture for the mechanism $\mc{M}_{\phi^n}^n(\theta)$, (b) CNN actor architecture for the mechanism $\mc{M}_{\phi^{n-1}}^{n-1}(\theta_{-i})$, (c) CNN actor architecture for the agent deviation policy $\pi'_{\phi^{n-1}, i}(\theta_i)$}
        \label{fig:mlp_actors}
\end{figure}

\begin{figure}[H]
     \centering
     \begin{subfigure}[b]{0.2\textwidth}
         \centering
         \includegraphics[width=\textwidth]{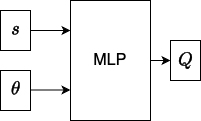}
         \caption{$\mc{M}_{\phi^n}^n(\theta)$}
         \label{fig:main_critic_mlp}
     \end{subfigure}
     \hfill
     \begin{subfigure}[b]{0.2\textwidth}
         \centering
         \includegraphics[width=\textwidth]{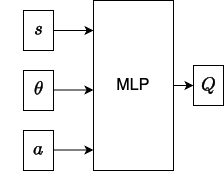}
         \caption{$\mc{M}_{\phi^{n-1}}^{n-1}(\theta_{-i})$}
         \label{fig:critic_nopart_multi_mlp}
     \end{subfigure}
     \hfill
     \begin{subfigure}[b]{0.2\textwidth}
         \centering
         \includegraphics[width=\textwidth]{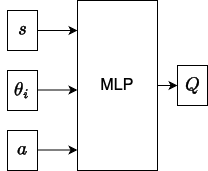}
         \caption{$\pi'_{\phi^{n-1}, i}(\theta_i)$}
         \label{fig:critic_nopart_single_mlp}
     \end{subfigure}
        \caption{(a) MLP critic architecture for $Q^n_{\psi^n}(\mc{M}^n_{\phi^n}(\theta),\theta)$, (b) MLP critic architecture for $Q^{n-1}_{\psi^{n-1}, -i}(\mc{M}^{n-1}_{\phi^{n-1}, -i}(\theta_{-i}),\theta)$, (c) MLP critic architecture for $Q^{n-1}_{\psi^{n-1}, i}(\pi'_{\phi', i}, \mc{M}^{n-1}_{\phi^{n-1}, -i}(\theta_{-i}),\theta_i)$}
        \label{fig:critic_nopart_single_mlp}
\end{figure}

\begin{figure}[H]
     \centering
     \begin{subfigure}[b]{0.3\textwidth}
         \centering
         \includegraphics[width=\textwidth]{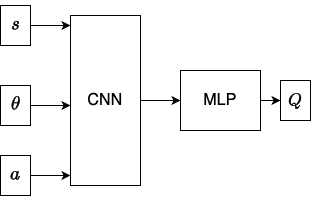}
         \caption{$\mc{M}_{\phi^n}^n(\theta)$}
         \label{fig:main_critic_cnn}
     \end{subfigure}
     \hfill
     \begin{subfigure}[b]{0.3\textwidth}
         \centering
         \includegraphics[width=\textwidth]{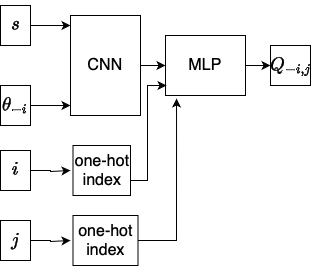}
         \caption{$\mc{M}_{\phi^{n-1}}^{n-1}(\theta_{-i})$}
         \label{fig:critic_nopart_multi_cnn}
     \end{subfigure}
     \hfill
     \begin{subfigure}[b]{0.3\textwidth}
         \centering
         \includegraphics[width=\textwidth]{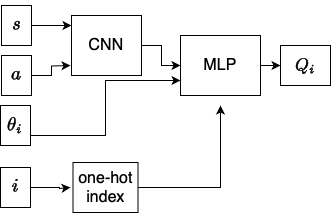}
         \caption{$\pi'_{\phi^{n-1}, i}(\theta_i)$}
         \label{fig:critic_nopart_single_cnn}
     \end{subfigure}
        \caption{(a) CNN critic architecture for $Q^n_{\psi^n}(\mc{M}^n_{\phi^n}(\theta),\theta)$, (b) CNN critic architecture for $Q^{n-1}_{\psi^{n-1}, -i}(\mc{M}^{n-1}_{\phi^{n-1}, -i}(\theta_{-i}),\theta)$, (c) CNN critic architecture for $Q^{n-1}_{\psi^{n-1}, i}(\pi'_{\phi', i}, \mc{M}^{n-1}_{\phi^{n-1}, -i}(\theta_{-i}),\theta_i)$}
        \label{fig:critic_nopart_single_cnn}
\end{figure}

\begin{figure}[H]
     \centering
     \begin{subfigure}[b]{0.25\textwidth}
         \centering
         \includegraphics[width=\textwidth]{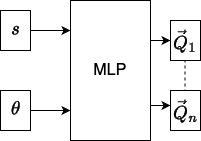}
         \caption{$Q^n_{\psi^n}(\theta)$}
         \label{fig:main_critic_vdn_mlp}
     \end{subfigure}
     \hfill
     \begin{subfigure}[b]{0.25\textwidth}
         \centering
         \includegraphics[width=\textwidth]{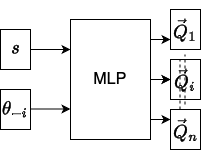}
         \caption{$Q^{n-1}_{\psi^{n-1}, -i}(\theta_{-i})$}
         \label{fig:nopart_multi_vdn_mlp}
     \end{subfigure}
     \hfill
     \begin{subfigure}[b]{0.25\textwidth}
         \centering
         \includegraphics[width=\textwidth]{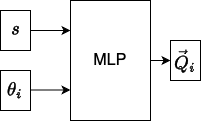}
         \caption{$Q^{n-1}_{\psi^{n-1}, i}(\theta_i)$}
         \label{fig:nopart_single_vdn_mlp}
     \end{subfigure}
        \caption{(a) MLP critic architecture for the mechanism Q-network $Q^n_{\psi^n}(\theta)$, (b) MLP critic architecture for the mechanism Q-network $Q^{n-1}_{\psi^{n-1}, -i}(\theta_{-i})$, (c) MLP critic architecture for the agent deviation Q-network $Q^{n-1}_{\psi^{n-1}, i}(\theta_i)$}
        \label{fig:mlp_critics_vdn}
\end{figure}

\begin{figure}[H]
     \centering
     \begin{subfigure}[b]{0.25\textwidth}
         \centering
         \includegraphics[width=\textwidth]{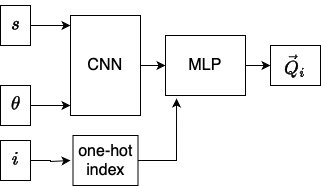}
         \caption{$Q^n_{\psi^n}(\theta)$}
         \label{fig:main_critic_vdn_cnn}
     \end{subfigure}
     \hfill
     \begin{subfigure}[b]{0.25\textwidth}
         \centering
         \includegraphics[width=\textwidth]{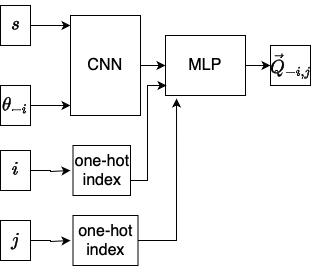}
         \caption{$Q^{n-1}_{\psi^{n-1}, -i}(\theta_{-i})$}
         \label{fig:nopart_multi_vdn_cnn}
     \end{subfigure}
     \hfill
     \begin{subfigure}[b]{0.25\textwidth}
         \centering
         \includegraphics[width=\textwidth]{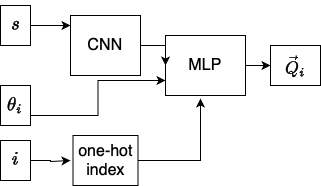}
         \caption{$Q^{n-1}_{\psi^{n-1}, i}(\theta_i)$}
         \label{fig:nopart_single_vdn_cnn}
     \end{subfigure}
        \caption{(a) CNN critic architecture for the mechanism Q-network $Q^n_{\psi^n}(\theta)$, (b) CNN critic architecture for the mechanism Q-network $Q^{n-1}_{\psi^{n-1}, -i}(\theta_{-i})$, (c) CNN critic architecture for the agent deviation Q-network $Q^{n-1}_{\psi^{n-1}, i}(\theta_i)$}
        \label{fig:mlp_critics_vdn}
\end{figure}

\begin{figure}[H]
     \centering
     \begin{subfigure}[b]{0.45\textwidth}
         \centering
         \includegraphics[width=.5\textwidth]{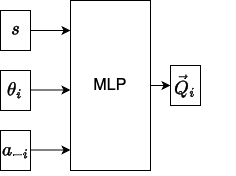}
         \caption{$Q_{\psi^{Q}, i}(\mc{M}^n_{\phi^n, -i}(\theta),\theta_i)$, $V_{\psi^{V}, i}(\mc{M}^n_{\phi^n}(\theta),\theta_i)$}
         \label{fig:ind_critic_mlp}
     \end{subfigure}
     \hfill
     \begin{subfigure}[b]{0.45\textwidth}
         \centering
         \includegraphics[width=.5\textwidth]{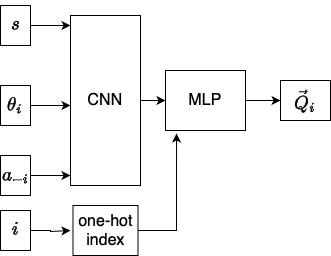}
         \caption{$Q_{\psi^{Q}, i}(\mc{M}^n_{\phi^n, -i}(\theta),\theta_i)$, $V_{\psi^{V}, i}(\mc{M}^n_{\phi^n}(\theta),\theta_i)$}
         \label{fig:ind_critic_cnn}
     \end{subfigure}
        \caption{(a) MLP critic architecture for the Q-networks $Q_{\psi^{Q}, i}(\mc{M}^n_{\phi^n, -i}(\theta),\theta_i)$, $V_{\psi^{V}, i}(\mc{M}^n_{\phi^n}(\theta),\theta_i)$, (b) CNN critic architecture for the Q-networks $Q_{\psi^{Q}, i}(\mc{M}^n_{\phi^n, -i}(\theta),\theta_i)$, $V_{\psi^{V}, i}(\mc{M}^n_{\phi^n}(\theta),\theta_i)$}
        \label{fig:mlp_critics_vdn}
\end{figure}

\subsection{MA-TD3 Details}

For MA-TD3, MADDPG \cite{maddpg} is used as base with full communication of state and types $\theta$ using Gumbel Softmax \cite{gumbelsoftmax} to ensure differentiability for discrete action space, and modifications for TD3 \cite{td3} are included accordingly. We employ clipped double-q learning, use 0.1 entropy on the policy network for target policy smoothing, and do the delayed policy updates with a period of 5.

\subsection{Evaluation Details}

For evaluation, IC and IR losses are estimated with the converged critic neural networks at the end of the training. Since the mechanism policy of ReMBo is changing throughout the training, policies from  $\mc{M}_{\phi^n}^n(\theta)$ and $\mc{M}_{\phi^{n-1}}^{n-1}(\theta_{-i})$ are saved to the evaluation buffer at evaluation steps to be used at the end of the training to compute IC and IR losses.

For all evaluations, 5 seeds are used. The results are reported with the thick lines (straight or dashed) denoting median and the error bars denoting 0.25'th and 0.75'th quantiles. The results are smoothed with Pandas exponentially weighted moving average with an $\alpha = 0.25$.

For Matrix Games, the evaluation is done every 100 steps for 20 episodes per evaluation. For Lane-Changing Game and Congestion Games, the evaluation is done every 500 steps for 40 episodes per evaluation. 
\newpage
\section{Additional Experiment Information}

\subsection{Repeated Matrix Games}

The payoff matrices for the Chicken Game with Types and Stag Hunt with Preferences is given below. For these games, the IC and IR loss are computed performing a brute force search for the losses of all possible mechanism policies. The losses are averaged over time steps. For these games, MLP based neural network architectures are used, where the state, types, and actions are fed as a flattened vector.

\newcommand{\specialcell}[2][c]{%
\begin{tabular}[#1]{@{}c@{}}#2\end{tabular}}

\begin{table}[H]
\caption{Chicken Game with Types - Payoffs}\label{tab:chicken}
    \begin{subtable}{0.5\textwidth}
    \centering
    \begin{tabular}{p{1.5cm} |cc}
        & \specialcell{Chicken} & \specialcell{Dare} \\
        \hline
        \specialcell{Chicken} & (2, 2) & (1, 3) \\
        \hline
        \specialcell{Dare} & (3, 1) & (0, 0) \\
        \end{tabular} 
        \caption{RA - RA Payoff Table}
    \end{subtable}%
    \begin{subtable}{0.5\textwidth}
    \centering
        \begin{tabular}{p{1.5cm} |cc}
        & \specialcell{Chicken} & \specialcell{Dare} \\
        \hline
        \specialcell{Chicken} & (2, 1) & (1, 3) \\
        \hline
        \specialcell{Dare} & (3, 0) & (0, 1) \\
        \end{tabular}
        \caption{RA - RT Payoff Table} 
    \end{subtable}
    \medskip

    \begin{subtable}{0.5\textwidth}
    \centering
    \begin{tabular}{p{1.5cm} |cc}
        & \specialcell{Chicken} & \specialcell{Dare} \\
        \hline
        \specialcell{Chicken} & (1, 2) & (0, 3) \\
        \hline
        \specialcell{Dare} & (3, 1) & (1, 0) \\

        \end{tabular} 
        \caption{RT - RA Payoff Table}
    \end{subtable}%
    \begin{subtable}{0.5\textwidth}
    \centering
        \begin{tabular}{p{1.5cm} |cc}
        & \specialcell{Chicken} & \specialcell{Dare} \\
        \hline
        \specialcell{Chicken} & (1, 1) & (0, 3) \\
        \hline
        \specialcell{Dare} & (3, 0) & (1, 1) \\
        \end{tabular}
        \caption{RT - RT Payoff Table} 
    \end{subtable}

\end{table}

\begin{table}[H]
\caption{Stag Hunt with Preferences - Payoffs}\label{tab:stag}
    \begin{subtable}{0.5\textwidth}
    \centering
    \begin{tabular}{p{1.5cm} |cc}
        & \specialcell{Stag} & \specialcell{Rabbit} \\
        \hline
        \specialcell{Stag} & (3, 3) & (0, 1) \\
        \hline
        \specialcell{Rabbit} & (1, 0) & (1, 1) \\
        \end{tabular} 
        \caption{S - S Payoff Table}
    \end{subtable}%
    \begin{subtable}{0.5\textwidth}
    \centering
        \begin{tabular}{p{1.5cm} |cc}
        & \specialcell{Stag} & \specialcell{Rabbit} \\
        \hline
        \specialcell{Stag} & (3, 2) & (0, 3) \\
        \hline
        \specialcell{Rabbit} & (1, 0) & (1, 3) \\
        \end{tabular}
        \caption{S - R Payoff Table} 
    \end{subtable}
    \medskip

    \begin{subtable}{0.5\textwidth}
    \centering
    \begin{tabular}{p{1.5cm} |cc}
        & \specialcell{Stag} & \specialcell{Rabbit} \\
        \hline
        \specialcell{Stag} & (2, 3) & (0, 1) \\
        \hline
        \specialcell{Rabbit} & (3, 0) & (3, 1) \\

        \end{tabular} 
        \caption{R - S Payoff Table}
    \end{subtable}%
    \begin{subtable}{0.5\textwidth}
    \centering
        \begin{tabular}{p{1.5cm} |cc}
        & \specialcell{Stag} & \specialcell{Rabbit} \\
        \hline
        \specialcell{Stag} & (2, 2) & (0, 3) \\
        \hline
        \specialcell{Rabbit} & (3, 0) & (3, 3) \\
        \end{tabular}
        \caption{R - R Payoff Table} 
    \end{subtable}

\end{table}

\begin{figure}[H]
    \centering
        \includegraphics[width=.87\textwidth]{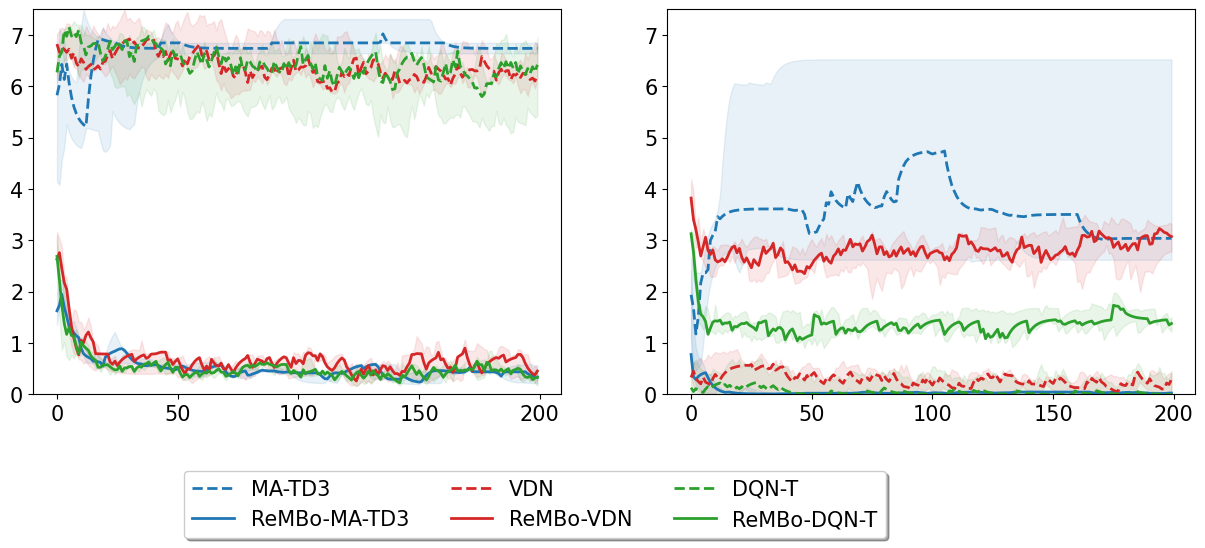}\\
        \includegraphics[width=.87\textwidth]{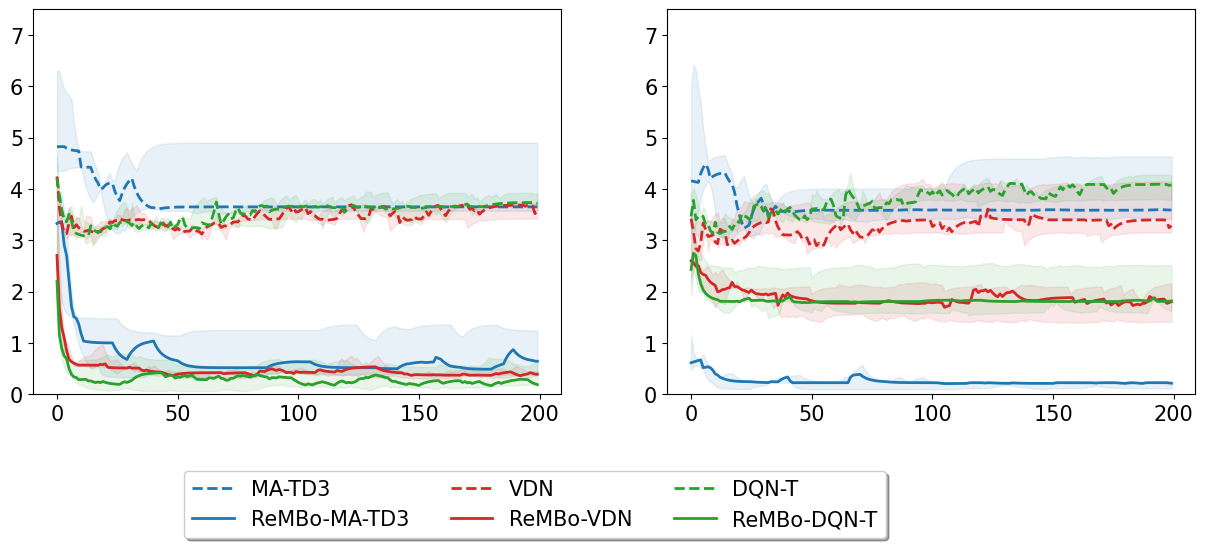}
    \caption{Results for \emph{Chicken} (top) and \emph{Stag Hunt} (bottom).  Left: Estimated IC deviation, right: Estimated IR deviation.}\label{fig:est_chicken_res}
\end{figure}




\subsection{Lane-Changing Game Model}


\begin{wrapfigure}{l}{0.25\textwidth}
\includegraphics[width=0.5\linewidth]{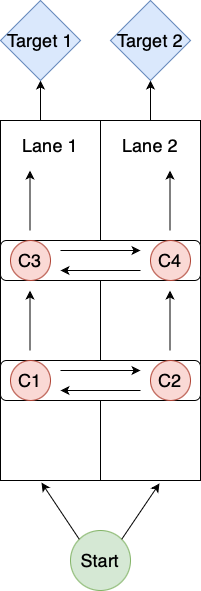} 
\caption{Lane-Changing Game}
\label{fig:lane}
\end{wrapfigure}

The game has a total $N$ number of drivers, out of $k$ number of drivers entering the game every step. Agents have a target $\theta \in \{\text{Target 1, Target 2}\}$ they have to reach which is private information to them. The agents that enter the game start from the node "Start" and pick 1 of 2 lanes. Lane 2 has 3 times the delay of Lane 1. There are two positions at each lane to change lanes which divide the road into 3 parts. Agents are affected by the traffic at their own parts of the lane they use, or the parts ahead of them of the lane they use. Agents can only get a positive reward of "20" (target reaching reward) if they reach their own target. Changing a lane has a cost to agents that change the lane themselves, or that are affected by the other agents that change the lane. For example, any lane change between C1 and C2 affect the agents that are at Start, C1, and C2. Any lane change between C3 and C4 affect the agents that are at Start, C1, C3, C3 or C4. Lane utilization is computed for 3 part of the lane at each lane, a total of 6 lane utilization values for both lanes. Lane utilization for each part is computed as the sum of agents using that part of the lane and the parts of the lanes ahead at the same lane. The delay cost for an agent is defined as:

\begin{align*}
    \left(\left(\text{lane\_utilization(agent)} + \text{affecting\_lane\_changing\_agents(agent)}\right) / k\right) * \text{lane\_coefficient}
\end{align*}

Reward of an agent per step is the delay cost at last step + target reaching reward.

Lane-Changing Game uses the CNN based neural network architectures. The state, types, and actions are fed in 2D shape in case of full communication, where each row has the relevant data per agent.

\subsection{Congestion Game Models}

Congestion Games use the CNN based neural network architectures. The state, types, and actions are fed in 2D shape in case of full communication, where each row has the relevant data per agent.

\begin{figure}[H]
     \centering
     \begin{subfigure}[b]{0.45\textwidth}
         \centering
         \includegraphics[width=.7\textwidth]{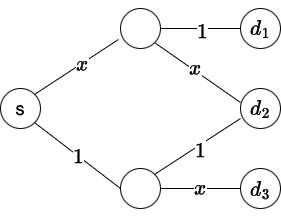}
         \label{fig:3type}
     \end{subfigure}
     \hfill
     \begin{subfigure}[b]{0.45\textwidth}
         \centering
         \includegraphics[width=.7\textwidth]{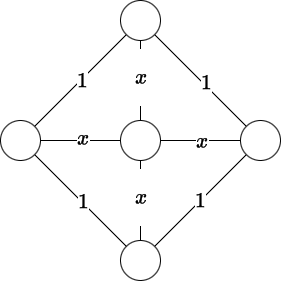}
         \label{fig:intersection}
     \end{subfigure}
        \caption{Left: 3-Destination Congestion Game. Right: Intersection Congestion Game.}
        \label{fig:congmodels}
\end{figure}

\section{Hyperparameters}\label{app:hyper}

For the ReMBo loss, we rescale the coefficients given in Equation \ref{E:mdproblem_loss} such that $\alpha_0 = 1$ and $\alpha_1$ and $\alpha_2$ are as given in the tables below.

For all experiments, epsilon greedy exploration is used. For Matrix Games, the decay schedule is from 0.8 to 0 with a decay rate of 0.999 each episode. For Lane-Changing Game the decay schedule is from 0.95 to  0.15 with a decay rate of 0.999 each episode (0.15 + $\epsilon$). For Congestion Games, the decay schedule is from 0.8 to 0.15 with a decay rate of 0.999 each episode (0.15 + $\epsilon$).

\begin{table}[H]
\caption{Hyperparameters}\label{hyperparam}
\begin{subtable}{0.5\textwidth}
    \centering
    \begin{tabular}{c|c}
        Hyperparameter & Value \\
        \hline
        $\gamma$ & 0.99 \\
        $lr\_actor$ & 1e-4 \\
        $lr\_critic$ & 1e-3 \\
        $\tau$ & 0.01 \\
        batch size & 32 \\
        training buffer size & 20000 \\
        gumbel temperature & 1 \\
        hidden layer 1 size & 64 \\
        hidden layer 2 size & 64 \\
        $\alpha_{1_{MA-TD3}}$ & 50 \\
        $\alpha_{2_{MA-TD3}}$ & 50 \\
        $\alpha_{1_{VDN}}$ & 50 \\
        $\alpha_{2_{VDN}}$ & 50 \\
        $\alpha_{1_{DQN-T}}$ & 50 \\
        $\alpha_{2_{DQN-T}}$ & 50 \\
        \end{tabular} 
        \caption{Matrix Games Hyperparameters}
        \end{subtable}
        \medskip
        \begin{subtable}{0.5\textwidth}
        \begin{tabular}{c|c}
        Hyperparameter & Value \\
        \hline
        $\gamma$ & 0.99 \\
        $lr\_actor$ & 1e-4 \\
        $lr\_critic$ & 5e-4 \\
        $\tau$ & 0.01 \\
        batch size & 16 \\
        training buffer size & 1000000 \\
        gumbel temperature & 1 \\
        hidden layer 1 size & 64 \\
        hidden layer 2 size & 64 \\
        number of filters & 6 \\
        kernel size & (3, 3) \\
        stride & (1, 1) \\
        $\alpha_{1_{MA-TD3}}$ & 50 \\
        $\alpha_{2_{MA-TD3}}$ & 50 \\
        $\alpha_{1_{VDN}}$ & 50 \\
        $\alpha_{2_{VDN}}$ & 50 \\
        $\alpha_{1_{DQN-T}}$ & 50 \\
        $\alpha_{2_{DQN-T}}$ & 50 \\
        \end{tabular} 
        \caption{Lane-Changing Game Hyperparameters}
        \end{subtable}

        \begin{subtable}{0.5\textwidth}
        \begin{tabular}{c|c}
        Hyperparameter & Value \\
        \hline
        $\gamma$ & 0.99 \\
        $lr\_actor$ & 1e-4 \\
        $lr\_critic$ & 1e-3 \\
        $\tau$ & 0.01 \\
        batch size & 32 \\
        training buffer size & 1000000 \\
        gumbel temperature & 1 \\
        hidden layer 1 size & 64 \\
        hidden layer 2 size & 64 \\
        number of filters & 6 \\
        kernel size & (3, 3) \\
        stride & (1, 1) \\
        $\alpha_{1_{MA-TD3}}$ & 20 \\
        $\alpha_{2_{MA-TD3}}$ & 20 \\
        $\alpha_{1_{VDN}}$ & 100 \\
        $\alpha_{2_{VDN}}$ & 100 \\
        $\alpha_{1_{DQN-T}}$ & 100 \\
        $\alpha_{2_{DQN-T}}$ & 100 \\
        \end{tabular} 
        \caption{3-Destination Congestion Game Hyperparameters}
        \end{subtable}
        \medskip
        \begin{subtable}{0.5\textwidth}
        \begin{tabular}{c|c}
        Hyperparameter & Value \\
        \hline
        $\gamma$ & 0.99 \\
        $lr\_actor$ & 1e-4 \\
        $lr\_critic$ & 1e-3 \\
        $\tau$ & 0.01 \\
        batch size & 32 \\
        training buffer size & 1000000 \\
        gumbel temperature & 1 \\
        hidden layer 1 size & 64 \\
        hidden layer 2 size & 64 \\
        number of filters & 6 \\
        kernel size & (3, 3) \\
        stride & (1, 1) \\
        $\alpha_{1_{MA-TD3}}$ & 20 \\
        $\alpha_{2_{MA-TD3}}$ & 20 \\
        $\alpha_{1_{VDN}}$ & 100 \\
        $\alpha_{2_{VDN}}$ & 100 \\
        $\alpha_{1_{DQN-T}}$ & 100 \\
        $\alpha_{2_{DQN-T}}$ & 100 \\
        \end{tabular} 
        \caption{Intersection Congestion Game Hyperparameters}
        \end{subtable}

\end{table}

\section{Compute Resources}\label{app:comp}
A collection of NVIDIA A40 and NVIDIA RTX A6000 servers are used for the experiments.



\end{document}